\documentstyle[sprocl,epsf]{article}
\newcommand{\Nf}{N_{\!f}} 
\newcommand{\MSbar}{\overline{\mbox{MS}}} 
\newcommand{\Dslash}{D \! \! \! \! /} 
\newcommand{\kslash}{k \! \! \! /}

\begin{document}
\begin{flushright} 
{\bf {LTH 382}} 
\end{flushright} 
\vspace{0.4cm} 
\title{RENORMALIZATION GROUP FUNCTIONS OF QCD IN LARGE $N_{\! f}$ \footnote{
Invited talk presented at 3rd International Conference on the Renormalization 
Group, JINR, Dubna, Russia, 26-31st August, 1996.} 
} 
\author{ J.A. GRACEY }
\address{Department of Mathematical Sciences, University of Liverpool, P.O. Box
147, Liverpool, L69 3BX, United Kingdom} 
\maketitle\abstracts{We review the application of the critical point large 
$\Nf$ self-consistency method to QCD. In particular we derive the $O(1/\Nf)$ 
$d$-dimensional critical exponents whose $\epsilon$-expansion determines the
perturbative coefficients in $\MSbar$ of the field dimensions, $\beta$-function
and various twist-$2$ operators which occur in the operator product expansion 
of deep inelastic scattering.} 

\section{Introduction} 
The renormalization group equation, (RGE), plays an important role in 
comparing predictions made in a quantum field theory with observations of 
nature. The fundamental ingredients in the RGE are the renormalization group 
functions. Since these are rarely known exactly even for the simplest of field
theories one has to be content with approximate perturbative solutions; the 
accuracy being dependent upon how many orders in the perturbative coupling 
constant one can compute the RGE functions. This is a highly technical and 
tedious exercise partly because the number of Feynman diagrams at even one 
loop can sometimes be excessive. Also the results depend on how one removes the
ultra-violet infinities. For theories which particle physicists are interested 
in such as quantum chromodynamics, (QCD), which is the gauge theory describing 
the strong interactions, most high order calculations of these functions are 
performed in the $\MSbar$ scheme.\cite{1}$^{\!-\,}$\cite{4} For instance, the 
$\beta$-function of QCD has been deduced at third order in this scheme. 
Recently information on various scattering amplitudes has been produced at the 
same level in an impressive set of papers.\cite{5} Due to the complexity of 
such calculations, having independent and alternative methods to check the high 
order structure of the RGE functions is important. 

One such method has been made available through the properties of the RGE in 
the neighbourhood of a fixed point which is defined to be a non-trivial zero of
the $\beta$-function. There it is known that the critical exponents which 
characterize the phase transition correspond to the functions of the RGE  
evaluated at the critical coupling. So if one can compute exponents directly 
then information on the RGE functions is obtainable.\cite{6} This has been 
achieved in impressive articles by Vasil'ev et al for the $O(N)$ $\sigma$ 
model.\cite{7} There critical exponents were determined in arbitrary dimensions
order by order in powers of $1/N$ when $N$ is large. Those results are in total
agreement with the $\epsilon$-expansion at the fixed point of the same 
exponents deduced explicitly at $5$-loops in $\MSbar$. Garnered by that success
it is therefore a worthwhile exercise to develop the $1/\Nf$ method for QCD, 
where $\Nf$ is the number of quark flavours, in relation to the present state 
of the art calculations. 

\section{Basic ideas} 
We recall the basic ideas for deducing arbitrary dimensional critical exponents
in the $1/\Nf$ expansion. First from the two loop $\beta$-function of QCD in 
$d$-dimensions,\cite{1}$^{\!,\,}$\cite{2} there is a fixed point at 
\begin{eqnarray} 
g_c &=& \frac{3\epsilon}{T(R)\Nf} + \frac{1}{4T^2(R)\Nf^2} \left[ \frac{}{} 
33C_2(G)\epsilon - \left( 27C_2(R) + 45C_2(G)\right) \epsilon^2 \right.
\nonumber \\ 
&&+~ \left. \left( \frac{99}{4}C_2(R) + \frac{237}{8}C_2(G) \right) 
\epsilon^3 + O(\epsilon^4) \right] + O \left( \frac{1}{\Nf^3} \right)
\end{eqnarray}
where $d$ $=$ $4$ $-$ $2\epsilon$. If, for example, a general RGE function 
takes the form 
\begin{equation} 
\gamma(g) ~=~ c_1g + (c_2\Nf+d_1)g^2 + (c_3\Nf^2+d_2\Nf+e_1)g^3 + O(g^4) 
\end{equation} 
where the coefficients $\{c_i, d_i, e_i \ldots \}$ are independent of $\Nf$, 
then the associated exponent at leading order in $1/\Nf$ is 
\begin{equation} 
\gamma(g_c) ~=~ \frac{1}{\Nf} \sum_{r=1}^\infty c_r [3\epsilon/T(R)]^r ~+~ 
O(1/\Nf^2) 
\end{equation} 
So provided $\gamma(g_c)$ can be computed directly in the large $\Nf$ limit its 
$\epsilon$-expansion gives the leading order sequence of coefficients 
$\{c_i\}$ of $\gamma(g)$. 

The exponents are defined with reference to the action of the theory one is 
interested in. For QCD this takes the form 
\begin{equation} 
L ~=~ i \bar{\psi}^{iI} \Dslash \psi^{iI} ~-~ \frac{(G^a_{\mu \nu})^2}{4e^2}   
\end{equation} 
where $\psi^{iI}$ is the quark field, $A_\mu^a$ is the gluon field, $D_\mu$ $=$ 
$\partial_\mu$ $+$ $T^a A^a_\mu$, $G^a_{\mu\nu}$ $=$ $\partial_\mu A^a_\nu$ 
$-$ $\partial_\nu A^a_\mu$ $+$ $f^{abc}A^b_\mu A^c_\nu$, $T^a_{IJ}$ is the 
generator of the colour group whose structure constants are $f^{abc}$, $1$ 
$\leq$ $i$ $\leq$ $\Nf$, $1$ $\leq$ $I$ $\leq$ $N_c$ and $1$ $\leq$ $a$ $\leq$ 
$(N^2_c$ $-$ $1)$. The canonical dimensions of the fields of Eq. 4 at $g_c$ are
defined by demanding that the action is dimensionless. The anomalous dimensions
are defined to be the extra portion of the full dimension of the field or 
operator and essentially are a measure of the effect of radiative corrections. 
For instance, in the scaling region where the propagators of Eq. 4 behave in 
the limit $k^2$ $\rightarrow$ $\infty$, as,\cite{8}  
\begin{equation}
\psi(k) ~\sim~ \frac{A\kslash}{(k^2)^{\mu-\alpha}} ~~,~~
A_{\mu\nu}(k) ~\sim~ \frac{B}{(k^2)^{\mu-\beta}}\left[ \eta_{\mu\nu}
- (1-b)\frac{k_\mu k_\nu}{k^2} \right] 
\end{equation} 
where $A$ and $B$ are momentum independent amplitudes and $b$ is the covariant
gauge parameter, we define  
\begin{equation} 
\alpha ~=~ \mu ~-~ 1 ~+~ \frac{1}{2} \eta ~~~,~~~ \beta ~=~ 1 ~-~ \eta ~-~ \chi 
\end{equation} 
with $d$ $=$ $2\mu$. Here $\chi$ is the dimension of the quark gluon vertex 
operator and $\eta$ is the quark anomalous dimension. Expressions for these 
anomalous dimensions are deduced from studying the scaling dimensions of the 
next to leading order corrections to the $2$ and $3$ point Green's function 
using Eq. 5.\cite{7} For an arbitrary gauge parameter, the leading order 
results are,\cite{8}  
\begin{eqnarray}
\eta &=& \frac{C_2(R)[(2\mu-1)(\mu-2)+\mu b]\eta^{\mbox{o}}_1} 
{(2\mu-1)(\mu-2)T(R)\Nf} \\ 
\eta \, + \, \chi &=& - \, \frac{C_2(G)[(2\mu-1)+b(\mu-1)] \eta^{\mbox{o}}_1} 
{2(2\mu-1)(\mu-2)T(R)\Nf} 
\end{eqnarray}
where $\eta^{\mbox{o}}_1$ $=$ $-$ $(2\mu-1)(2-\mu)\Gamma(2\mu)/[4\Gamma^2(\mu)
\Gamma(\mu+1)\Gamma(2-\mu)]$.  

In computing these results, which agree with $3$-loop perturbative calculations
in the Landau gauge,\cite{4} we made use of another well known feature of 
critical point theory. Ordinarily more than one model can be used to deduce 
exponents at a fixed point and such models are said to be in the same 
universality class. A well known example is the equivalence of the $O(N)$ 
$\sigma$ model and $O(N)$ $\phi^4$ theory in three dimensions. For the present
case QCD is equivalent$\,$\cite{9} at leading order in $1/\Nf$ to a non-abelian 
version of the Thirring model, (NATM), which is renormalizable in strictly two 
dimensions. Its lagrangian is  
\begin{equation} 
L ~=~ i \bar{\psi}^{iI} \Dslash \psi^{iI} ~-~ \frac{(A^a_\mu)^2}{2\lambda}   
\end{equation} 
where $\lambda$ is the coupling constant which is dimensionless in $2$
dimensions. Eliminating the auxiliary spin-$1$ field $A^a_\mu$ yields a 
$4$-fermi term. The benefit of using this model, Eq. 9, is that it has a 
simpler structure to Eq. 4 as the $3$ and $4$ point gluon self interactions are
absent. So one need only consider diagrams built with the quark gluon 
interaction. It was shown, though,\cite{9} that in the $1/\Nf$ limit the 
$4$-fermi model correctly reproduced the $3$ and $4$ point gluon Feynman rules 
in the approach to four dimensions. In other words with Eq. 9 the effect of the
$3$-point gluon interaction is contained in the graphs with a quark loop. This 
feature occurs implicitly in the calculations we report on later. Further in 
using a covariant gauge, ghost fields have to be included in each lagrangian 
but they give no contribution at leading order.  

\section{$\beta$-function} 
With this basic formalism the $O(1/\Nf)$ correction to the QCD 
$\beta$-function can be computed.\cite{10} Ordinarily this is the first step in
determining $O(1/\Nf^2)$ information as it will encode the next order 
correction to $g_c$ to all orders in $\epsilon$. To determine this 
we compute the related exponent $\omega$ $=$ $-$ $\beta^\prime(g_c)/2$. 
It is deduced from the last term of Eq. 4 which gives the scaling law 
\begin{equation} 
\omega ~=~ \eta ~+~ \chi ~+~ \chi_G 
\end{equation} 
where $\chi_G$ is the critical dimension of the composite operator $G$ $=$ 
$(G^a_{\mu\nu})^2$ when computed as an insertion in a Green's function in the
non-abelian Thirring model. For QED $\omega$ was originally deduced in $1/\Nf$ 
by explicitly performing the $\MSbar$ renormalization with an infinite chain of
electron bubbles.\cite{11} The extension to the non-abelian case is simpler in 
the critical approach. Three $2$-loop and one $3$-loop graphs need to be 
evaluated which are illustrated in Fig. 1. 
\begin{figure}[h]  
\vspace{0.5cm} 
\hspace{2.5cm} 
\epsfxsize=7cm 
\epsfbox{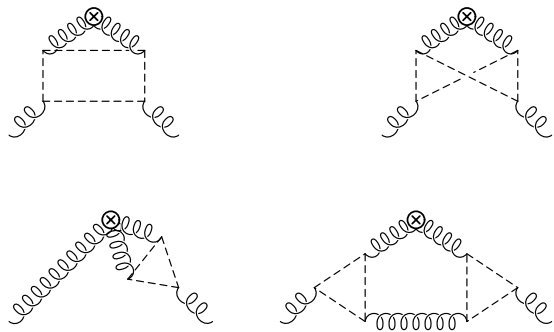} 
\caption{Graphs for $O(1/\Nf)$ contribution to $\omega$.} 
\end{figure} 
The first two graphs correspond to the QED sector, whilst the remaining
two would be absent by Furry's theorem in QED as their colour group factor is 
$C_2(G)$. Consequently, using the critical propagators we find  
\begin{eqnarray} 
\omega &=& (\mu-2) ~-~ \left[ (2\mu-3)(\mu-3)C_2(R) \right. \nonumber \\ 
&&~~~~ \left. -~ \frac{(4\mu^4 - 18\mu^3 + 44\mu^2 - 45\mu + 14)C_2(G)} 
{4(2\mu-1)(\mu-1)} \right] \frac{\eta^{\mbox{o}}_1}{T(R)\Nf}  
\end{eqnarray} 
The $\epsilon$-expansion of Eq. 11 correctly reproduces the $O(1/\Nf)$ 
coefficients of the $3$-loop $\MSbar$ 
$\beta$-function.\cite{1}$^{\!-\,}$\cite{4} With this agreement we can deduce 
several new higher order coefficients. Using the notation
\begin{equation} 
\beta(g) ~=~ (d-4)g ~+~ \left( \frac{2}{3} T(R) \Nf - \frac{11}{6}C_2(G) 
\right) g^2 ~+~ \sum_{r=2}^\infty a_r[T(R)\Nf]^{r-2} g^{r+1} 
\end{equation}  
for the large $\Nf$ leading order part of the $\beta$-function, 
then$\,$\cite{10}
\begin{eqnarray} 
a_4 &=& - ~ [154C_2(R) + 53C_2(G)]/3888 \nonumber \\ 
a_5 &=& [(288\zeta(3) + 214)C_2(R) + (480\zeta(3) - 229)C_2(G)]/31104 
\nonumber \\  
a_6 &=& [(864\zeta(4) - 1056\zeta(3) + 502)C_2(R) \nonumber \\ 
    && +~ (1440\zeta(4) - 1264\zeta(3) - 453)C_2(G)]/233280 \nonumber \\  
a_7 &=& [(3456\zeta(5) - 3168\zeta(4) - 2464\zeta(3) + 1206)C_2(R) \nonumber \\
    && +~ (5760\zeta(5) - 3792\zeta(4) - 848\zeta(3) - 885)C_2(G)]/1679616 
\end{eqnarray} 

\section{Twist-$2$ operators} 
With the impressive progress that has been made at $3$-loops in $\MSbar$ in the 
renormalization of the twist-$2$ operators of the operator product expansion 
used to understand processes in deep inelastic scattering$\,$\cite{5} it is 
important to have some large $\Nf$ results available for comparison. Similar to
the $\beta$-function calculation the critical exponents corresponding to the 
anomalous dimensions of such operators are deduced by inserting the operator 
into a Green's function in the NATM and determining the scaling behaviour of 
the integrals. The operators which we consider are, 
\begin{eqnarray}
{\cal O}^{\mu_1 \ldots \mu_n}_{\mbox{\footnotesize{ns}}} &=&  
i^{n-1} {\cal S} \bar{\psi}^I \gamma^{\mu_1} D^{\mu_2} \ldots D^{\mu_n} 
T^a_{IJ} \psi^J - \mbox{trace terms} \nonumber \\ 
{\cal O}^{\mu_1 \ldots \mu_n}_{\mbox{\footnotesize{s}}} &=&  
i^{n-1} {\cal S} \bar{\psi}^I \gamma^{\mu_1} D^{\mu_2} \ldots D^{\mu_n} 
\psi^I - \mbox{trace terms} \nonumber \\ 
{\cal O}^{\mu_1 \ldots \mu_n}_{\mbox{\footnotesize{g}}} &=&  
\frac{i^{n-2}}{2} {\cal S} \, \mbox{tr} \, G^{a \, \mu_1\nu} D^{\mu_2} 
\ldots D^{\mu_{n-1}} G^{a \, ~ \mu_n}_{~~\nu} - \mbox{trace terms} \nonumber 
\end{eqnarray}
where ${\cal S}$ denotes symmetrization on the Lorentz indices. 

For the fermionic twist-$2$ flavour nonsinglet and singlet operators, 
${\cal O}_{\mbox{\footnotesize{ns}}}$ and ${\cal O}_{\mbox{\footnotesize{s}}}$,
we deduce at leading order in $1/\Nf$ respectively$\,$\cite{12}  
\begin{eqnarray}
\eta^{(n)}_{{\footnotesize{\mbox{ns}}}} &=&  
\frac{2C_2(R)(\mu-1)^2\eta^{\mbox{o}}_1}{(2\mu-1)(\mu-2)T(R)\Nf} 
\left[ \frac{(n-1)(2\mu+n-2)}{(\mu+n-1)(\mu+n-2)} ~+~ \frac{2\mu}{(\mu-1)} 
\Psi(n) \right] \nonumber \\  
\eta_{{\footnotesize{\mbox{s}}}}^{(n)} 
&=& \frac{(\mu-1)C_2(R)\eta^{\mbox{o}}_1}{(2\mu-1)
(\mu-2)T(R)\Nf} \left[ \frac{2(\mu-1)(n-1)(2\mu+n-2)}{(\mu+n-1)(\mu+n-2)} 
{}~+~ 4\mu\Psi(n) \right. \nonumber \\
&&- \, \left. \mu\Gamma(n-1)[(n^2+n+2\mu-2)^2 + 2(\mu-2)(n(n-1)(2\mu-3+2n) 
\right. \nonumber \\ 
&& \left. +~ 2(\mu-1+n))]\Gamma(2\mu)/[(\mu+n-1)(\mu+n-2)\Gamma(2\mu-1+n)] 
\frac{}{} \! \right]  
\end{eqnarray}
where $n$ is the operator moment, $\Psi(n)$ $=$ $\psi(\mu-1+n)$ $-$ $\psi(\mu)$
and $\psi(x)$ is the logarithmic derivative of the $\Gamma$-function. One 
feature of the singlet sector is that the operators do not mix since the 
gluonic and fermionic operators have different canonical dimensions at $g_c$. 
By contrast in the perturbative calculation there is mixing and one has to 
compute a matrix of anomalous dimensions. To compare the $\epsilon$ expansion 
of Eq. 14 with perturbative results one realises that in the large $\Nf$ 
calculation the result contained in Eq. 14 is in fact the anomalous dimension 
of the predominantly fermionic eigenoperator of the perturbative mixing matrix.
Therefore if one computes the eigenvalues of the mixing
matrix$\,$\cite{5}$^{\!,\,}$\cite{13} and evaluates them at $g_c$ the 
coefficients of both $\epsilon$ expansions ought to be in agreement. We record 
this occurs exactly at the $3$-loop level at leading order in $1/\Nf$.  

More explicitly we present the $n$-dependence of the coefficient $c_3$, in the 
notation of Eq. 2, of both the nonsinglet and singlet leading order large $\Nf$
part of the anomalous dimensions. Having the explicit dependence is important 
since the inverse Mellin transform of the anomalous dimensions with respect to
$n$ determine the Altarelli Parisi splitting functions. These are a function of
the conjugate variable, $x$, which is the momentum fraction carried by the 
partons contained in the nucleons, and are in effect a measure of the 
probability that a parton fragments into other partons. First, we have for the 
nonsinglet case,\cite{12}  
\begin{eqnarray}
c_3^{\footnotesize{\mbox{ns}}} &=& \frac{2}{9}S_3(n) ~-~ 
\frac{10}{27}S_2(n) ~-~ \frac{2}{27}S_1(n) ~+~ \frac{17}{72} \nonumber \\
&&-~ \frac{[12n^4+2n^3-12n^2-2n+3]}{27n^3(n+1)^3} 
\end{eqnarray} 
where $S_l(n)$ $=$ $\sum_{r=1}^n 1/r^l$. To compare with the results of the 
explicit $3$-loop $\MSbar$ calculation for the first few moments,\cite{5} we 
have evaluated Eq. 15 for various $n$ and presented them in Table 1. 
{\begin{table}[ht] 
\hspace{3cm} 
{ \begin{tabular}{c|r}  
$n$ & $c_3^{\footnotesize{\mbox{ns}}}$ \\ 
\hline
& \\
$2$ & $- \, \frac{28}{243}$ \\
& \\
$4$ & $ - \, \frac{384277}{1944000}$ \\
& \\
$6$ & $ - \, \frac{80347571}{333396000}$ \\
& \\
$8$ & $ - \, \frac{38920977797}{144027072000}$ \\  
& \\
$10$ & $ - \, \frac{27995901056887}{95850016416000}$ \\
& \\
$12$ & $ - \, \frac{65155853387858071}{210582486065952000}$ \\
& \\
$14$ & $ - \, \frac{68167166257767019}{210582486065952000}$ \\
& \\
$16$ & $ - \, \frac{5559466349834573157251}{16553468064672354816000}$ \\
& \\
$18$ & $ - \, \frac{19664013779117250232266617}{56770118727793840841472000}$ \\
& \\
$20$ & $ - \, \frac{6730392290450520870012467}
{18923372909264613613824000}$ \\
& \\
$22$ & $ - \, \frac{16759806821032136669044226177}
{46048135637404510767879321600}$ \\ 
\end{tabular} } 
\vspace{0.2cm}
{\caption{Values of $c_3^{\footnotesize{\mbox{ns}}}$ for various $n$.} }
\end{table}}
For the singlet sector we can deduce the $n$-dependence of the $3$-loop 
coefficient of the anomalous dimension of the predominantly fermionic
eigenoperator. It is,\cite{12}  
\begin{eqnarray} 
c_3^{\footnotesize{\mbox{s}}} &=& \frac{2}{9}S_3(n) ~-~ \frac{10}{27}S_2(n) ~-~
\frac{2}{27}S_1(n) ~+~ \frac{17}{72} \nonumber \\ 
&&-~ \frac{2(n^2+n+2)^2[S_2(n)+S^2_1(n)]}{3n^2(n+2)(n+1)^2(n-1)} \nonumber \\
&&-~ 2S_1(n)[16n^7+74n^6+181n^5+266n^4+269n^3+230n^2 \nonumber \\ 
&&~~~~~~~~~~~~~~~ + \, 44n-24]/[9(n+2)^2(n+1)^3(n-1)n^3] \nonumber \\
&&-~ [100n^{10}+682n^9+2079n^8+3377n^7+3389n^6 \nonumber \\ 
&&~~~~~ + \, 3545n^5+3130n^4+118n^3-940n^2-72n \nonumber \\ 
&&~~~~~ + \, 144]/[27(n+2)^3(n+1)^4n^4(n-1)]  
\end{eqnarray}
Similar to $c_3^{\footnotesize{\mbox{ns}}}$ we have evaluated Eq. 16 for low 
moments and presented the results in Table 2. These are in exact agreement with
the first four moments of the explicit three loop $\MSbar$ results after 
diagonalizing the mixing matrix \cite{5} and extracting the leading order large
$\Nf$ piece corresponding to the dimension of the predominantly fermionic 
eigenoperator. 
{\begin{table}[h] 
\hspace{3cm} 
{ \begin{tabular}{c|r}  
$n$ & $c_3^{\footnotesize{\mbox{s}}}$ \\ 
\hline 
& \\ 
$2$ & $0$ \\
& \\ 
$4$ & $ - \, \frac{121259}{720000}$ \\
& \\ 
$6$ & $ - \, \frac{3166907}{13891500}$ \\
& \\ 
$8$ & $ - \, \frac{1328467729}{5038848000}$ \\
& \\ 
$10$ & $ - \, \frac{304337312935261}{1054350180576000}$ \\  
& \\ 
$12$ & $ - \, \frac{842357166098254633}{2737572318857376000}$ \\  
& \\ 
$14$ & $ - \, \frac{42512567719680559}{131614053791220000}$ \\  
& \\ 
$16$ & $ - \, \frac{755896148277147625515451}{2251271656795440254976000}$ \\  
& \\ 
$18$ & $ - \, \frac{1121815282809553973842772849}
                       {3235896767484248927963904000}$ \\  
& \\ 
$20$ & $ - \, \frac{78640886458671664340562623}
                       {220772683941420492161280000}$ \\  
& \\ 
$22$ & $ - \, \frac{4248342909129791924572989157741}
                       {11650178316263341224273468364800}$ \\  
\end{tabular} } 
\vspace{0.2cm}
{\caption{Values of $c_3^{\footnotesize{\mbox{s}}}$ for various $n$.} }
\end{table}}

Aside from agreeing with explicit perturbative results up to three loops, 
there are several other checks on the exponents arising from general 
principles. First, as the operators are physical their anomalous dimensions 
are gauge independent. We have therefore computed Eq. 14 with a non-zero 
covariant gauge parameter $b$ and observed its cancellation in assemblying the 
contributions from the relevant Feynman diagrams in each exponent. Second, for 
certain values of $n$ the corresponding operators reduce to conserved physical 
currents. Provided the conservation of these currents is not spoiled by an 
anomaly then their anomalous dimensions must be zero to all orders in 
perturbation theory. For the nonsinglet sector the $n$ $=$ $1$ case relates to 
charge conservation, whilst the singlet operator with $n$ $=$ $2$ corresponds 
to the energy momentum tensor. Therefore for both these respective values the 
critical exponents of Eq. 14 must vanish. It is an easy exercise to verify 
this. Indeed the zero entry for $n$ $=$ $2$ in Table 2 is a reflection of this 
general result in the three loop case.  

\section{Conclusions} 
The critical renormalization group ideas$\,$\cite{7} have proved useful in 
giving some insight into the structure of the $\MSbar$ perturbative 
coefficients at higher orders in QCD. Although we have concentrated on the four
dimensional theory the results have been expressed as functions of $d$. 
Therefore we can also obtain information on the three dimensional model. For 
example, from Eq. 11  
\begin{equation} 
\omega ~=~ - \, \frac{1}{2} ~-~ \frac{10C_2(G)}{3\pi^2T(R)N_{\! f}} ~+~ 
O \left( \frac{1}{\Nf^2} \right)
\end{equation} 
Higher order $1/\Nf$ calculations are possible too. For instance, in the 
abelian sector the dimension of the mass operator, $\bar{\psi} \psi$, is 
available in $d$-dimensions. So when $d$ $=$ $3$ the gauge independent electron
mass anomalous dimension is,\cite{14} 
\begin{equation} 
\gamma_m(g_c) ~=~ -~ \frac{32}{3\pi^2\Nf} ~-~ 
\frac{64[3\pi^2 - 28]}{9\pi^4\Nf^2} ~+~ O \left( \frac{1}{\Nf^3} \right)
\end{equation}
Such results will be useful for comparing with numerical results for the same 
quantity computed by other methods. Indeed exponents which are known to similar
orders in other models like the $O(N)$ $4$-fermi model and evaluated for low 
$N$ have been in good agreement with lattice results.\cite{15} 

\section*{Acknowledgement} 
This work was carried out in part through a {\sc pparc} Advanced Fellowship.

\end{document}